\documentclass{aa}
\usepackage{amssymb}
\usepackage{graphicx}
\usepackage{hyperref}

\def\asec{\ifmmode ^{\prime\prime}\else$^{\prime\prime}$\fi}
\def\etal{{et\,al. }}
\def\msun{M$_{\odot}$}
\def\degs{\ifmmode ^{\circ}\else$^{\circ}$\fi}
\def\amin{\ifmmode ^{\prime}\else$^{\prime}$\fi}
\def\asec{\ifmmode ^{\prime\prime}\else$^{\prime\prime}$\fi}
\def\farcs{\hbox{$.\!\!^{\prime\prime}$}}  

\def\degs{\ifmmode ^{\circ}\else$^{\circ}$\fi}
\def\amin{\ifmmode ^{\prime}\else$^{\prime}$\fi}

\def\wl{~\lambda~}
\def\wll{\lambda\lambda~}
\def\kms{\rm ~km~s^{-1}}
\def\gcm{\rm ~g~cm^{-3}}
\def\EE#1{\times 10^{#1}}

\def\lsim{\!\!\!\phantom{\le}\smash{\buildrel{}\over
  {\lower2.5dd\hbox{$\buildrel{\lower2dd\hbox{$\displaystyle<$}}\over
                               \sim$}}}\,\,}

\def\gsim{\!\!\!\phantom{\ge}\smash{\buildrel{}\over
  {\lower2.5dd\hbox{$\buildrel{\lower2dd\hbox{$\displaystyle>$}}\over
                               \sim$}}}\,\,}

\begin{document}

\title{Three-dimensional modeling of Type Ia supernovae - The power of
late time spectra
\thanks{Based  on observations collected at the European
Southern Observatory,
Paranal, Chile (ESO Programmes 67.D-0134 and 69.D-0193).}}

\author{
Cecilia~Kozma\inst{1},
Claes~Fransson\inst{1},
Wolfgang~Hillebrandt\inst{2},
Claudia~Travaglio\inst{3,2}, 
Jesper~Sollerman\inst{1},
Martin~Reinecke\inst{2},
Friedrich~Konrad~R\"opke\inst{2},
Jason Spyromilio\inst{4}}

\institute{
\inst{1} Stockholm Observatory, AlbaNova, Department of Astronomy, 106 91 Stockholm, Sweden 
\\
\inst{2} Max-Planck-Institut f\"{u}r Astrophysik,
Karl-Schwarzschild-Strasse 1, D-85741 Garching, Germany \\
\inst{3} Istituto Nazionale di Astrofisica (INAF) - Osservatorio Astronomico
di Torino, via Osservatorio 20, 10025 Pino Torinese (Torino), Italy\\
\inst{4} European Southern Observatory,
Karl-Schwarzschild-Strasse 2, D-85748 Garching, Germany
} 

\date{Received --- ?, Accepted --- ?} 
\authorrunning{Kozma \etal}
\titlerunning{3D modeling} 
\offprints{cecilia@astro.su.se}

\abstract{Late time synthetic spectra of Type Ia supernovae, based on
three-dimensional deflagration models, are presented.  We mainly focus
on one model, ``c3\_3d\_256\_10s'', for which the hydrodynamics 
(R\"opke 2005)
and
nucleosynthesis 
(Travaglio \etal 2004) 
was calculated 
up to the homologous phase of the
explosion. Other models with different ignition conditions and
different resolution are also briefly discussed.  
The synthetic spectra are compared to observed late time spectra.
We find that while
the model spectra after 300 to 500 days show a good agreement with the
observed Fe~II-III features, they also show too strong O~I and C~I
lines compared to the observed late time spectra. 
The oxygen and carbon emission originates from the low-velocity
unburned material in the central regions of these models. To
get agreement between the models and observations we find that
only a small mass of unburned material may be left in the center after the
explosion. This may be a problem for 
pure deflagration models, although improved initial conditions, as well
as higher resolution decrease the discrepancy. The relative intensity
from the different ionization stages of iron is sensitive to the
density of the emitting iron-rich material. We find that clumping,
with the presence of low density regions, is needed to reproduce the
observed iron emission, especially in the range between 4000 and 6000 \AA.
Both temperature and ionization depend sensitively on density, 
abundances and radioactive content. This work therefore illustrates 
the importance of
including the inhomogeneous nature of realistic three-dimensional
explosion models.  We briefly discuss the implications of the spectral
modeling for the nature of the explosion.
 }

\maketitle

\section{Introduction}
During the past few years Type Ia supernovae (SNe Ia) have become the main
tool to measure the expansion rate of the Universe with the surprising result
that it is presently in a phase of accelerated expansion. The importance of SNe
Ia as distance indicators comes from the fact that they are bright and can be
observed at large distances. Moreover, the homogeneity of their observable
properties, such as light curves and spectra, makes it reasonable to
calibrate the light curves in such a way that they become ``standard candles''.
In fact, empirical recipes to perform the calibration have been very
successful.  Today, by decreasing the statistical errors by the 
rapidly increasing number of observed SNe Ia at
intermediate and high redshifts, attempts are made to constrain also
the equation of state of the ``dark
energy'' that is thought to be the cause of the acceleration. 

However, in any such studies it is important to understand the 
systematic errors, which are here likely to dominate.
The intrinsic dispersion, i.e. originating
from properties of the explosions cannot be controlled by observing a
large number of SNe Ia. We therefore here compare realistic models to 
detailed observations to get a better understanding of the physics
leading to the homogeneity of the observed properties, and the inherent 
uncertainties.

The presently favored model for a Type Ia supernova is the thermonuclear
explosion and disruption of an accreting carbon-oxygen white dwarf with a mass
close to the Chandrasekhar limit. Guided by the presence of intermediate-mass
elements in the observed spectra these models assume that thermonuclear
burning begins as a (subsonic) deflagration, which may or may not change into
a (supersonic) detonation at later times. Recently, there has been great progress
in  the modeling of such explosions by means of combustion-hydrodynamics codes. 
In particular, three-dimensional deflagration calculations of increasing
complexity have been performed by 
a number of groups,
Reinecke \etal (2002a, 2002b), Gamezo \etal (2003) and 
Garcia-Senz \& Bravo (2005).  
Although there are
rather large differences in the ways turbulent thermonuclear flames are
modeled by the 
groups, they arrive at similar conclusions, namely that
the chemical structure of the supernova should be highly inhomogeneous. In
particular, they find substantial amounts of unburned carbon and oxygen close
to the center. Whether this is a result of inadequate modelling of the nuclear
burning, or is a real feature can best be decided by comparing model spectra
with observations.

In this paper we demonstrate the possibility of testing the multidimensional
hydrodynamics and nucleosynthesis calculations by studying the late time
spectra. The emission at late times, i.e. later than $\sim$ 100 days, is
predominantly emerging from the central parts of the supernova, where the
effects of the explosion are most pronounced.
Therefore, by comparing model spectra, based on different explosion models, to
observations one can draw conclusions about the explosive nucleosynthesis,
hydrodynamic mixing, the amount of $^{56}$Ni formed, and the total energy of
the explosion. In contrast, early spectra are more sensitive to the chemical
structure of the high velocity outer regions. The two phases therefore
nicely complement each other.  
The input models used in this study come from the
calculations by Travaglio \etal (2004), who have made a detailed post
processing nucleosynthesis of the hydrodynamic models by Reinecke \etal
(2002a, 2002b) and  R\"{o}pke (2005). The hydrodynamic code only had a 
minimal nucleosynthesis
network included, sufficient for a good approximation
of the thermonuclear energy release. 

In section \ref{modeling} we first summarize the three-dimensional
hydrodynamical models and the nucleosynthesis we use as input to our spectral
code. Then we discuss the late-time spectral modeling. Our
results
are given in section \ref{results}.  In section \ref{discussion} we discuss
our findings, and what we can learn from late time spectra. Finally in section
\ref{summary} we summarize our results.

\section{Modeling}
\label{modeling}
\subsection{3D hydrodynamical modeling}

Our input models for the spectral-synthesis code come from the
multidimensional hydrodynamic calculations of Reinecke \etal (2002a, 2002b).
In order to save computer time with still sufficient numerical resolution,
only an octant of the star was simulated and mirror symmetry was assumed at
the inner boundaries. This artificial symmetry causes some problems with the
shapes of the emission lines of the nebular spectra, as will be discussed
later.  Recently R\"{o}pke (2005) has evolved one of their three-dimensional
explosion models up to 10 seconds, at which time the assumption of homologous
expansion is valid. 

The simulations of Reinecke \etal (2002a, 2002b) attempt to model the nuclear
burning and hydrodynamics from first principles, as far as possible. The
nuclear burning is modeled as a subsonic, turbulent deflagration. Because the
length scales vary from sub-mm scales to dimensions comparable to the radius
of the white dwarf, some simplifying assumptions about the physics of the
flame front on small length scales have to be invoked.  The first assumption
is that since the flamefront cannot be numerically resolved it is replaced by
a sharp discontinuity between ``fuel'' and ``ashes''. This discontinuity is
then modeled by a level-set function. The remaining unknown is the normal
velocity of the level set, i.e. the turbulent flame speed. Since in the
flamelet regime of turbulent combustion (valid for Type Ia supernovae down to
densities of around 10$^7$ g cm$^{-3}$) the flame velocity is independent of
the physics on microscopic scales, it can be determined from a sub-grid scale
model of the unresolved turbulent velocity fluctuations. These, in turn,
result from a Kolmogorov-like turbulent cascade with energy fed in by
large-scale hydrodynamic instabilities, predominantly the Rayleigh-Taylor
(buoyancy-driven) instability. The important parameters determining the
evolution of the explosion are the central density and the composition of the
white dwarf. All simulations used in the present study started with a central
density of $2.9\EE9 \gcm$ and equal carbon and oxygen mass fractions.
 
The numerical experiments in Reinecke \etal (2002a, 2002b) show that the way
the explosion is initiated can have large effects on the outcome, especially
on the explosion energy and the $^{56}$Ni mass.  This is likely to be
determined by the evolution leading up to the ignition. According to
Garcia-Senz \& Woosley (1995) possible ignition conditions could be a couple
of floating ``blobs'' of burning material accelerated to a fraction of the
sound velocity near the center of the star. Alternatively, large scale
convective motions could lead to ignition in extended regions at the center
(H\"oflich \& Stein 2002), or a multi-point ignition with an exponentially
increasing and asymmetric number of ignition points out to approximately 
150 -- 200 km off-center (Woosley, Wunsch \& Kuhlen 2004). 

In the simulations by Reinecke \etal the effects of the ignition were
explored by assuming different ignition topologies. In particular, in
one set of models the explosion is initiated at the center, while in a
second set of calculations the explosion starts in a finite number of
randomly distributed, floating bubbles. The result of this is that
the more complex the initial conditions are, the more $^{56}$Ni seems
to form and the stronger the explosion energy. This is not too
surprising because, in general, a more complex geometry has a larger
surface area, increasing the burning rate and leading to a higher
energy. Also, the amount of unburned carbon and oxygen that is dredged-down
to the center between the rising Rayleigh-Taylor blobs depends on
the geometry of the ignition region, but this effect has not yet
been fully explored.

Finally, the numerical resolution is important, mostly for the nucleosynthesis
but, to a lesser extent also for the kinetic energy of the explosion. This is
especially important for the mass of $^{56}$Ni produced and also the amount of
unburned carbon and oxygen near the center of the star, and both are crucial parameters
for the calculations of synthetic spectra. The reason can easily be
understood. For instance, the number of initial burning blobs is limited by
the number of grid points. Low resolution means few blobs with rather large
sizes, in contrast to the few kilometers expected from analytical models
(Woosley, Wunsch \& Kuhlen 2004). These large blobs rise fast and burn little
fuel, leaving behind lots of unburned gas. In contrast, high resolution allows
for more realistic initial conditions and thus for more burning early on in
the explosion. Similar arguments hold for other initial conditions. The further
evolution, however, is much less affected because once turbulence has
developed the subgrid-scale model takes care of resolution
effects. Resolution studies underway (R\"opke \& Hillebrandt 2005c) seem to
indicate that even computations with (static) grids of 512$^3$ mesh points may
not describe the early evolution correctly as far as the nucleosynthesis
yields are concerned. We will come back to this question later.
 
In the present paper we will use three different 3D models, referred to as
b30\_3d\_768, b5\_3d\_256, and c3\_3d\_256. 
In Table \ref{modtab} we summarize the main characteristics of
the different hydrodynamical models studied here. 
The label c3 refers to central
burning, while b30 and b5 refer to models, where the burning starts in 30 and
5 blobs, respectively. The last number refers to the resolution of the model,
in theses cases $768^3$ or $256^3$ grid points, respectively. While the
b5\_3d\_256, and c3\_3d\_256 are from Reinecke \etal (2002b), the high
resolution model b30\_3d\_768 is discussed in Travaglio \etal (2004).
From what we discussed in the previous paragraph it is obvious that the
nuclear abundances obtained from the models with low numerical resolution
need to be considered with caution. 

The calculations by Reinecke \etal (2002b) follow the explosion up to 1.2 --
1.5 seconds.  After that time the outermost layers are 'escaping' their fixed
Eulerian grid. However, a problem for any spectral modeling is that the ejecta
have at that time not yet reached homologous expansion, but the velocities and
densities are still affected by pressure and gravity.  This problem has
recently been solved by R\"{o}pke (2005), who calculated the hydrodynamics for
the c3\_3d\_256 model by Reinecke \etal (2002b) up to 10 seconds after the
explosion, using a moving grid. We will in this paper refer to this model as
c3\_3d\_256\_10s. At 10 seconds the ejecta are indeed moving homologously, and
have reached their final velocities and densities.  In this paper we will
mainly concentrate on the c3\_3d\_256\_10s model, and just briefly discuss the
models b30\_3d\_768, c3\_3d\_256 and b5\_3d\_256 with respect to the initial
conditions and resolution.

The kinetic energies for the models are given in Table~\ref{modtab}.
These energies all refer to the end of the simulations, i.e. at 1.2 -- 1.5 s.
However, as shown by R\"{o}pke (2005), the kinetic energy does not change much
between 1.5 s and 10 s. The maximum velocity for the tracer particles (and the
supernova ejecta) in the c3\_3d\_256\_10s model is roughly $\sim 12800 \kms$.

\subsection{Nucleosynthesis calculations}

\noindent

Travaglio \etal (2004) have calculated the nucleosynthesis based on the
hydrodynamical models in Reinecke \etal (2002a, 2002b). 
The nucleosynthesis calculations
for the multidimensional explosion models are discussed in detail in  
Travaglio \etal. Here we just summarize the main features.
For computational reasons the hydrodynamical models of the explosions
only include a few representative nuclei (${}^4$He, $^{12}$C, $^{16}$O,
$^{24}$Mg and $^{56}$Ni) in a simplified reaction
network, sufficient to calculate the nuclear energy release. This is,
however, not sufficient for more detailed spectral models, or
for a detailed comparison of the nucleosynthesis. Therefore, a
post-processing step with a full nucleosynthesis network was performed,
with the physical conditions of 'tracer particles' as input. 

The reason for introducing such tracer particles is the need to 
map the Eulerian hydrodynamics grid onto a Lagrangian grid for
the nucleosynthesis computations. Therefore in the hydrodynamics a
number of tracer particles are followed, recording the temperature 
and density as a function of time, sufficiently large to represent
the matter of the exploding star. Technically this is done by subdividing 
the star into a Cartesian grid of ($27^3 = 19683$) cells, equidistant in 
the integrated mass, azimuthal angle $\phi$ and cos$\theta$ such that all 
cells contain the same mass: 1.39 \msun/19683 = $7\times10^{-5}$ \msun. 
A tracer particle is placed randomly in each of those grid cells and is 
then 'floating' with the expanding gas as the explosion proceeds.  
Since the hydrodynamics was computed in an octant only also the tracer 
particles cover one octant.

Knowing the thermal history of each particle, the nucleosynthesis for all
tracer particles in the 3D hydro model can be calculated. For the c3\_3d\_256
and b5\_3d\_256 low-resolution models the nucleosynthesis is calculated up to
1.5 seconds, while for the high resolution model b30\_3d\_768, the
temperatures and densities are extrapolated to 3 seconds. Freeze-out of
nuclear reactions was assumed to occur at a temperature of $1.5\times10^9$K in all
models, including the model c3\_3d\_256\_10s. The nuclear reaction network contains
383 nuclear species, and is based on Thielemann \etal (1996) and Iwamoto \etal
(1999). Weak interaction rates are updated from Langanke \& Martinez-Pinedo
(2000) and Martinez-Pinedo \etal (2000). 

\begin{figure*}[t]
\includegraphics[width=100mm,clip]{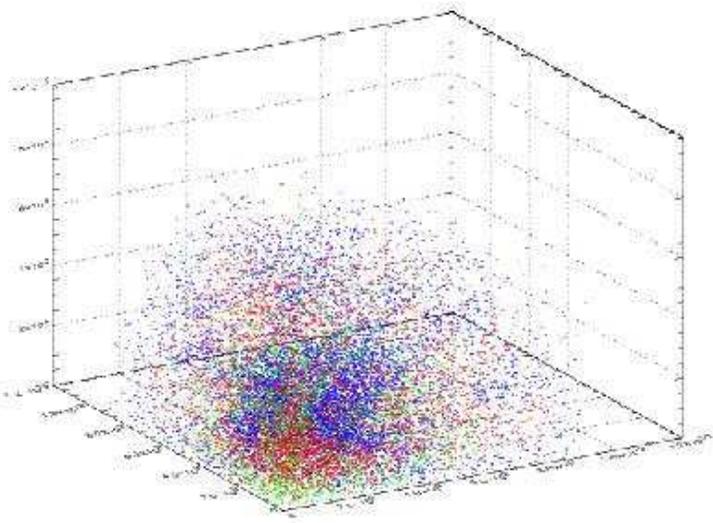}
\caption{
The positions of the Fe-rich (red), unburned (blue) and intermediate (green) 
tracer particles at 10 s in the c3\_3d\_256\_10s model.
Note the large amount of unburned particles in the central regions.
}
\label{posubb}
\end{figure*}

Initially, the tracer particles contain only $^{12}$C, $^{16}$O, and
$^{22}$Ne.  In all the models used in the present study for $\sim$ 30 to 40 \%
of the particles the temperatures are sufficiently high for nucleosynthesis to
transform the initial composition all the way up to iron group elements.  In
the remaining part, on the other hand, no or only incomplete burning occurs,
depending on the temperatures and densities.  
As the nucleosynthesis proceeds up to 10 seconds, the final abundances we use
in our nebular calculations are somewhat different from those in the
c3\_3d\_256 model at 1.5 seconds.  
This is mainly a consequence of the expanding computational grid in the model 
followed to 10 seconds, which reduces the flame resolution.
The mass of carbon in unburned particles is
0.31 \msun\ and of oxygen 0.35 \msun. Including also the partially processed
particles, the total mass of carbon and oxygen in the model is 0.34 and 0.42
\msun, respectively. The remaining particles have undergone nuclear
processing, resulting in intermediate mass elements and iron peak elements. We
refer to these as burned particles.  The total mass of $^{56}$Ni is in the
c3\_3d\_256\_10s model 0.28 \msun. This comparatively low mass may be affected by
the finite resolution, as we discuss below.  In Fig.~\ref{posubb} we show the
positions of the Fe-rich (red), unburned (blue), and intermediate (green) tracer 
particles in the c3\_3d\_256\_10s model.

\subsection{Late time spectral modeling}
\label{ltsm}

The modeling of late time spectra is based on the code described in detail in
Kozma \& Fransson (1998a). The code was originally used to model Type II
supernovae and has been applied to SN 1987A (Kozma \& Fransson 1998a; Kozma \&
Fransson 1998b). Since then the code has been improved in a number of ways,
and has also been extended to apply to Type Ia supernovae 
(Sollerman et al. 2004). We here therefore
summarize the main features and changes of the model.

The energy input to the ejecta at these late epochs is due to decays of
radioactive elements formed in the explosion.  In our calculations we include
decays of $^{56}$Ni to $^{56}$Co, and $^{56}$Fe, as well as decays of 
$^{57}$Ni and
$^{44}$Ti. We calculate the amount of gamma-ray and positron energy deposited
into heating, ionization, and excitation, by solving the Boltzmann equation,
as formulated by Spencer \& Fano (1954).  The relative fractions going into
these channels depend on the composition and degree of ionization.  In our
model the gamma-ray and positron deposition is therefore calculated for each 
tracer
particle separately.  Our treatment of the nonthermal deposition is described
in detail in Kozma \& Fransson (1992).

The temperature, ionization and level populations in each tracer particle
 are calculated in steady state for a specific time by solving the statistical
 equations, together with the energy balance. We find that steady 
state is a good approximation up to 
~500 days. At later times departures from steady state 
become increasingly important.

The following ions are included : H I-II, He I-III, C I-III, N I-II, O I-III,
Ne I-V, Na I-II, Mg I-III, Si I-III, S I-V, Ar I-V, Ca I-III, Fe I-V, Co I-V, Ni I-II.
Among these the following are treated as multilevel atoms, with the number of levels within
brackets: H I (20), He I (16), O I (13), Mg I (10), Si I (21), Ca II (6),
Fe I (121), Fe II (191), Fe III (112), Fe IV (45), Ni I (14), Ni II (18), Co II (108), 
Co III (87). To calculate the cooling and line emission from the remaining ions, the 
level populations are calculated in steady state, using two-, five-, or six-level atoms. 

Details and references for most of the atomic data used are given in 
Kozma \& Fransson (1998a). In addition to the atomic data given 
in this paper, the code has
been updated specifically to model Type Ia's (Sollerman et al. 2004).
In particular, higher ionization stages of especially the iron group
elements have been included.
The following charge transfer reactions important for the modeling of Type Ia's have been added :
Fe I + Fe II $\rightarrow$ Fe II + Fe I, Fe I + Fe III $\rightarrow$ Fe II + Fe II, and 
Fe I + Fe IV $\rightarrow$ Fe II + Fe III. Rate coefficients for these three reactions
are from Krstic, Stancil, \& Schultz (1997), as given by 
Liu \etal (1998).

We have improved the modeling of iron by updating the total
recombination rates and including rates to individual 
levels for Fe I (Nahar, Bautista, Pradhan 1997), Fe II (Nahar 1997), Fe III (Nahar 1996).

Both Co II and Co III are included as multilevel atoms. For Co II, with 108
levels, A-values are from Kurucz (1988) and Quinet (1998), and  
the energies are from Pickering \etal (1998). For Co III, with 87 levels, 
all data are from Kurucz (1988). 
Recombination rates to Co II and Co III are unknown. 
To estimate recombination
rates for these ions, we use the same values as for Fe II and Fe III.
We set the total recombination rates
to Co II and Co III  equal to the total recombination rates to Fe II and 
Fe III, and assume that the distribution to individual levels are 
similar to the individual levels in Fe II and Fe III.

Also collision strengths for the Co II and Co III transitions are 
unknown. 
As a rough approximation we set the collision strengths equal to
$\Omega = 0.1$ for forbidden 
transitions, and for the allowed transitions 
we use Van Regemorter's formula (van Regemorter 1962). As we show
below, the contribution of Co II-III is negligible at the epochs of
interest, and the lack of atomic data for these ions is therefore 
for this application not serious.

In all calculations presented in this paper we assume 
photoionization to be unimportant. As was argued in Kozma \& Fransson 
(1998b),  
the reason for this is that the high energy photons responsible
for the photoionization are absorbed 
and scattered by lines in the ejecta. 
Because the number of resonance transitions 
at these energies are extremely large
multiple resonance scattering, followed by branching into the optical,
will efficiently reduce the UV-field, responsible for the
photoionization.
In the optical range we do take fluoresence by the Ca II H and K lines 
explicitly into account 
in our calculation of the spectra. This is done by assuming that all emission
within $6000 \kms$, on the blue side of the H and K lines, to be
absorbed and re-emitted in the Ca II IR-triplet and the $\wll 7291, 7324$ 
lines. This is the dominant mechanism for the formation of these lines
(Kozma \& Fransson 1998b).
Other permitted lines are included, using the Sobolev
approximation to treat the scattering. We do, however, not include
fluoresence and scattering from line to line.
Branch \etal (2005) compare spectra modeled by the synthetic spectrum
code Synow, which include multiple scattering, but not any calculation
of the level populations, to observations of SN 1994D at nebular
phases.  From this they suggest that the effects of resonance scattering are
seen at $\sim$300 days in Ca II H, K and IR-triplet, and Na I. This is in agreement with the fact that these lines, as well as several Fe II lines, are optically thick also in our calculations.

\section{Results}
\label{results}
In this section we discuss the late-time spectra based on the c3\_3d\_256\_10s model
between 300 and 500 days. The reason for concentrating on these 
late epochs is that the optical depths in the optical and IR ranges are then
small enough not to severely affect the observed spectrum.
This is usually referred to as the nebular phase.
Observed spectra of high signal-to-noise (S/N) 
only exist up to $\sim$ 350 days, and our
comparison will therefore concentrate on this epoch.
From this we can then put constraints on the explosion models, such as 
the final amount of unburned material and density structure.

We will also briefly discuss the modeling
based on the b30\_3d\_768, b5\_3d\_256, and c3\_3d\_256 models.
However, for these three models the hydrodynamics and nucleosynthesis 
have only been calculated for the first 1.2 -- 1.5 seconds after explosion.
The ejecta have for these models not yet reached homologous 
expansion,  
and the densities in these models are significantly higher than
in the c3\_3d\_256\_10s model. As we will show, 
the physical conditions in the ejecta, and therefore the spectrum, are 
sensitive
to the density structure. Therefore, it is not possible
to draw any firm conclusions from late time modeling based on 
the 1.2s and 1.5s models. They might, however, indicate the effects
of different numerical resolution and
initial burning conditions on the late time spectra.

\begin{figure*}[t]
\includegraphics[width=130mm,clip,angle=-90]{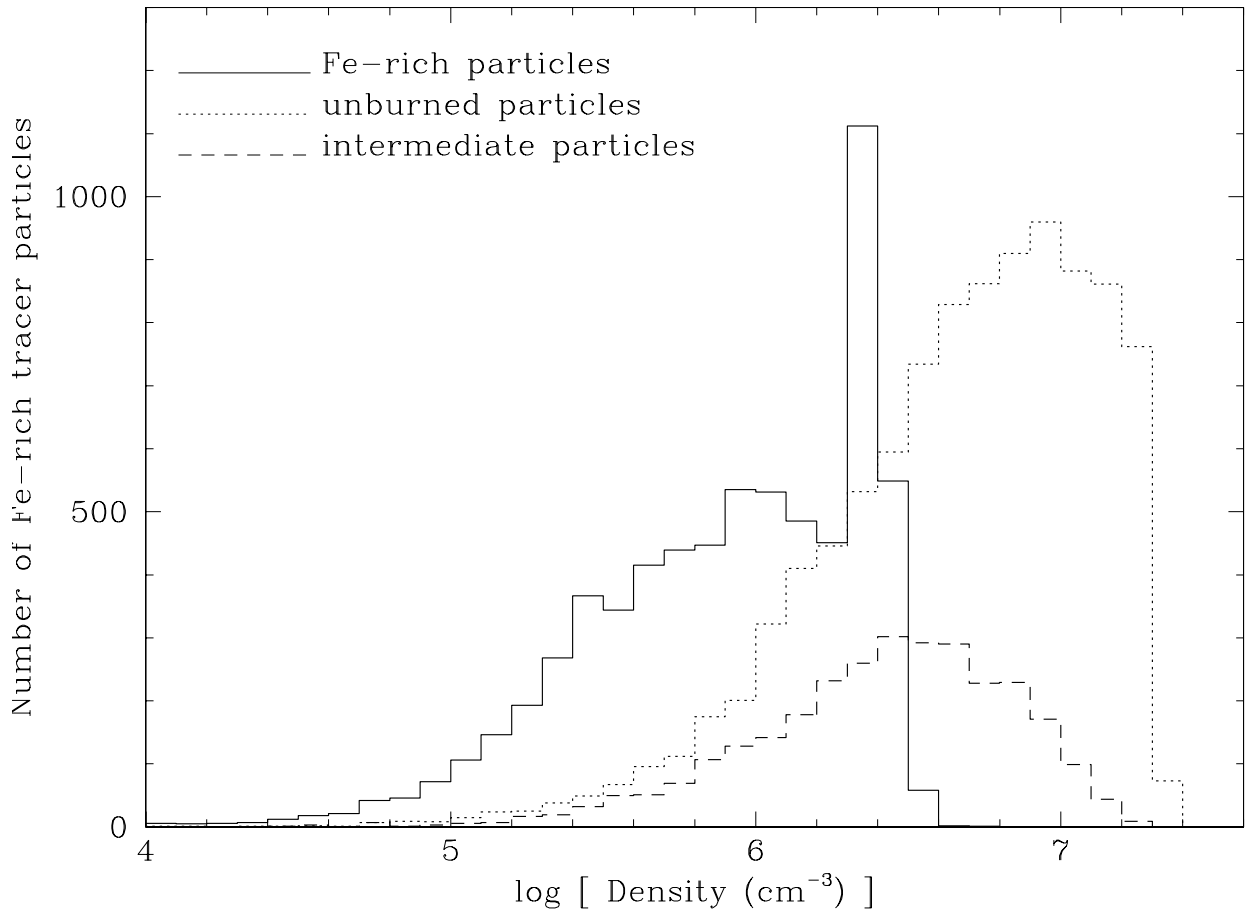}
\caption{
The distribution of number densities for the tracer particles 
in the c3\_3d\_256\_10s model at 300 days. 
}

\label{dendist}
\end{figure*}

Because of the importance of the density on the physical properties and
the emission, we have 
plotted the distribution of the number densities at 300 days for the
Fe-rich, intermediate mass elements, and unburned  
tracer particles in model c3\_3d\_256\_10s
in Fig.~\ref{dendist}. 
From this we see that there is a wide distribution in densities of the
Fe-rich particles from $\sim$ $3\times10^4$ to $3\times10^6$, with a
peak at $\sim 2\times10^6$ cm$^{-3}$. The densities of the unburned
particles are considerably higher, with a peak around 
$\sim 9\EE{6}$~cm$^{-3}$, while the particles with
intermediate mass elements are also intermediate in densities. This is
a natural consequence of the energy input due to the nuclear burning,
which expands mainly the regions where burning takes place. 

\subsection{Spectrum}
\label{spectrum}

In our models we calculate the ionization, temperature and level populations, 
and from these the  emissivity from each of the 19683
tracer particles for a given position (velocity), density, and
composition. From this we then construct a spectrum. The line transfer
is done in the Sobolev approximation.

As all tracer particles in these explosion models are located 
within one octant, the properties of each tracer particle have 
to be mirrored in the coordinate planes in order to model the full
ejecta. 
This simplified way of treating the geometry introduces
an artificial symmetry in the line profiles.
If we, for example,  view the ejecta along the positive z-axis, each of 
the particles in the model contributes to the emission at two distinct 
wavelengths. For each particle in the calculated octant,  
four particles, at $z_{particle} > 0$,  
contribute to the blue wing of a line, 
and the other four particles,  
at $z_{particle} < 0$, contribute to the red wing of the line. 
This gives rise to double-peaked line profiles, which is clearly an 
artifact of the assumption of symmetry with respect to the coordinate planes. 
This effect may be reduced by, for example, viewing the ejecta
from a different angle. We have tried this and, although the artificial
symmetry was reduced, it was still present.

To remove the effects of the assumed geometry we have recalculated
the spectra, assuming spherical symmetry for the emission. 
We divide the ejecta into 50 radial zones, and sum the emission from all
tracer particles within each radial interval. The line profiles and
spectra are then calculated from this spherically symmetric case. 
Note that this is only done {\it after} we have calculated the emission from 
each particle. The inhomogeneities of the model in terms of
density and composition, as well as the radial distribution, 
 are therefore preserved. 
All the spectra presented in the figures have been calculated
in this way.
In the future we hope
to discuss the detailed line profiles, based on a full $4\pi$ model
(R\"{o}pke \& Hillebrandt 2005b).

\begin{figure*}[t]
\includegraphics[width=180mm,clip]{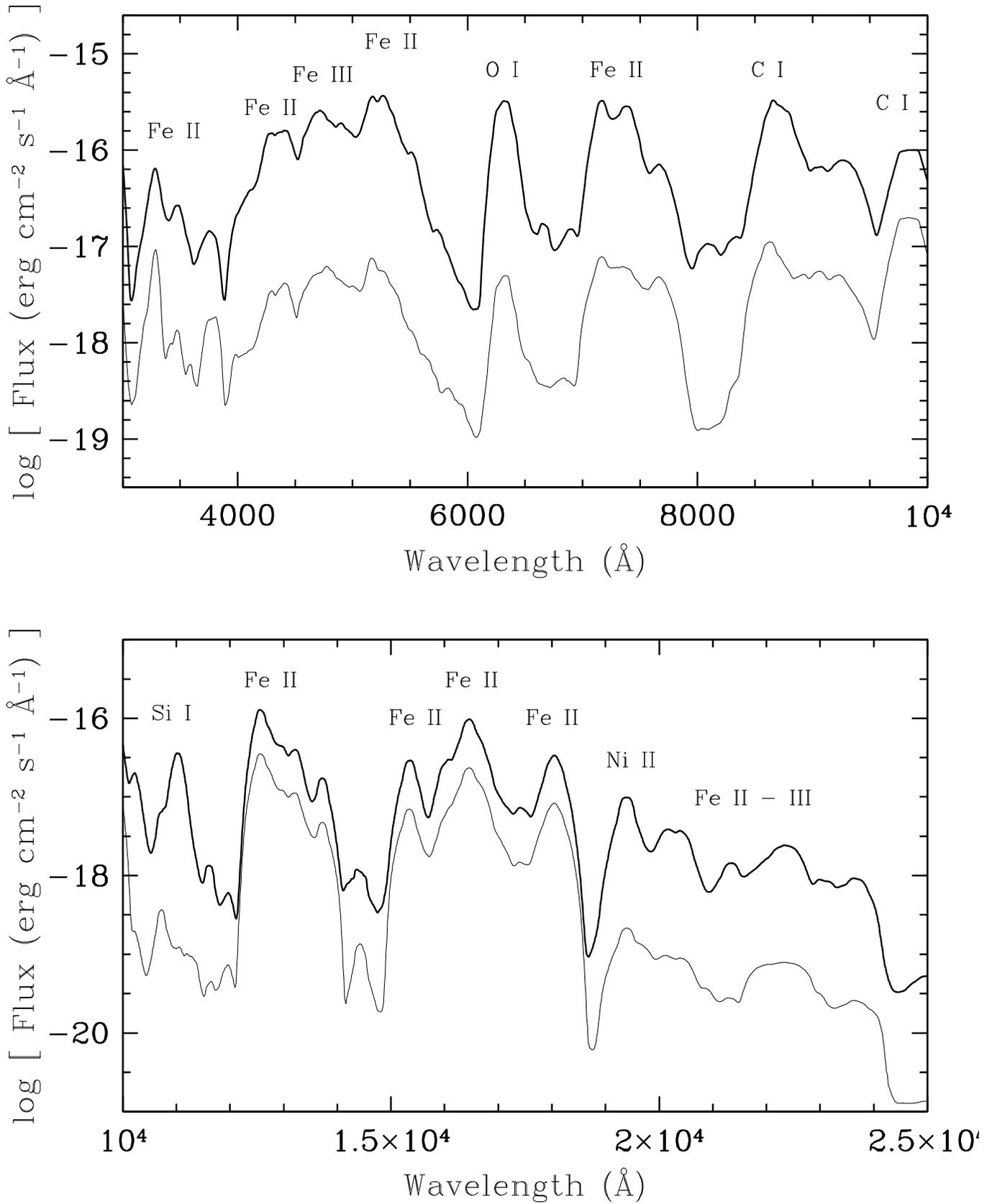}
\caption{Model spectra at 300 days (thick lines) and 500 days (thin lines) 
for the c3\_3d\_256\_10s model. The distance is assumed to be 10 Mpc.}
\label{spec35d}
\end{figure*}

\begin{figure*}[t]
\includegraphics[width=180mm,clip]{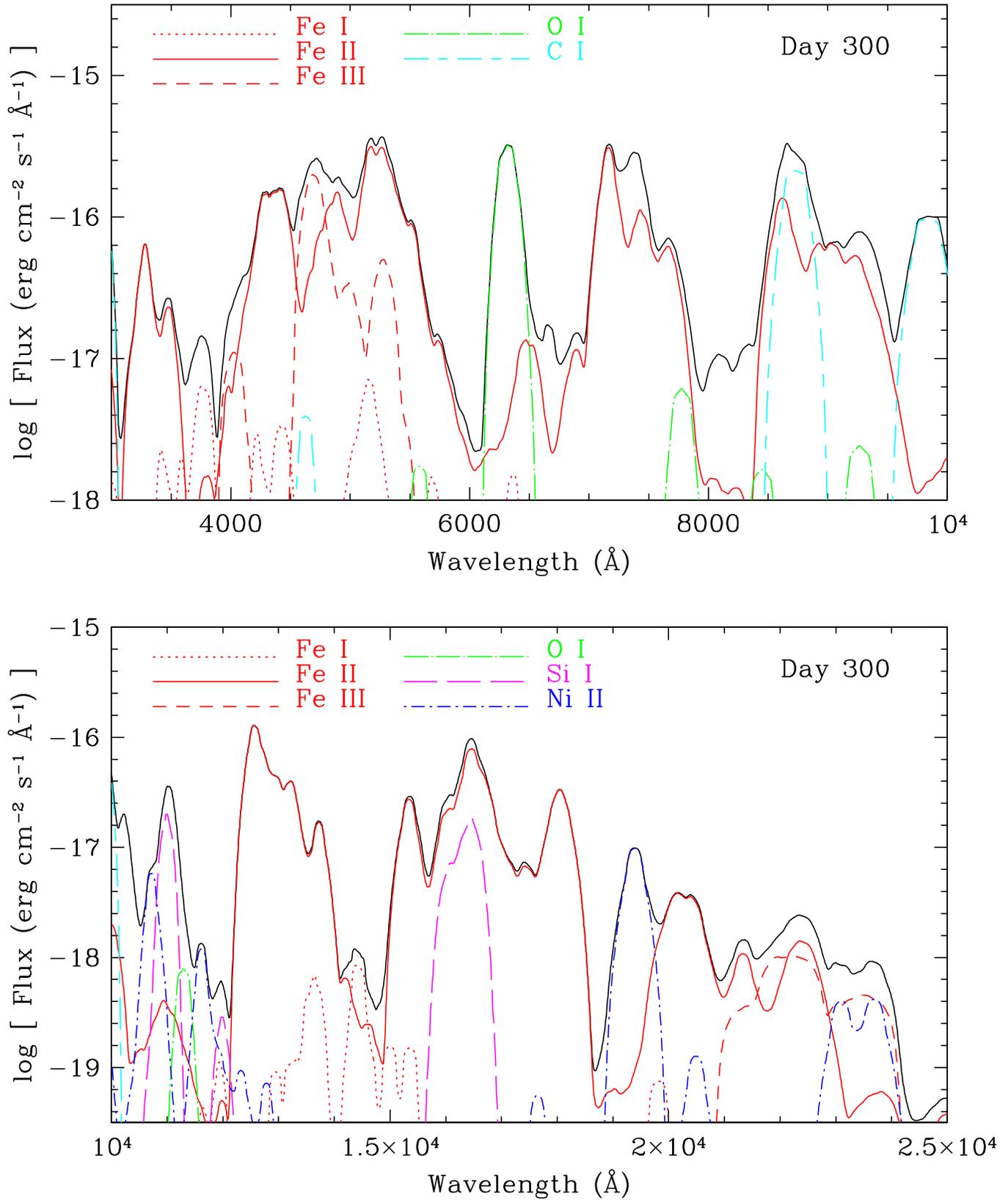}
\caption{The dominating contributions to the model spectra from c3\_3d\_256\_10s 
at 300 days. Here we show the contributions due to Fe~I, Fe~II, Fe~III, 
O~I, C~I, Si~I, and Ni~II.}
\label{specions513514}
\end{figure*}

In Fig.~\ref{spec35d} we show spectra for the  
two late epochs, 300 and 500 days past explosion,  
for the c3\_3d\_256\_10s model. 
The spectrum at 300 days is analyzed in more detail in 
Fig.~\ref{specions513514}, where we show the contributions
 from the lines of Fe I, Fe II, Fe III, Ni II, Si I, O I, and C I.

The main difference between the spectra at the two epochs is 
the decrease in ionization from day 300 to day 500,
as can be seen in the abundances of the iron ions.
At day 500 the emission
from Fe I has increased, and the emission from Fe II and III has decreased 
somewhat  
compared to the emission at 300 days. At both epochs, however, 
the dominant ion is Fe II. 

The most interesting results of these calculations are
the very strong lines from carbon and oxygen.
For both epochs [C~I] $\wll~9824, 9850$ and [O~I] $\wll~6300, 6364$ are
dominating the spectrum.
At 300 days also C~II] $\wl~2326$, C~I $\wll~2966, 2968$ and 
[C~I] $\wl~8727$ are strong, but they decrease in strength at day 500.
With regard to the UV lines, it should be noted that the spectrum 
below $\sim 3500 $ \AA\ is very uncertain due to the effects of
multiple scattering.

In the lower panel of Fig.~\ref{spec35d} we show the IR spectrum
between 1 and 2.5 $\mu$m. At these wavelengths there are no strong
oxygen or carbon lines. The spectrum is dominated by Fe~II, but also shows  
strong Si~I and Ni~II lines. The strongest lines are 
[Fe~II] $1.257 \mu$, $1.644 \mu$ and [Si~I] $1.099 \mu$.
The only intermediate mass element giving rise
to a significant emission feature in the models is silicon. As can
be seen in Fig.~\ref{spec35d}, the [Si~I] $1.099 \mu$ line is strong
in the 300 day spectrum, while it has almost disappeared at 500 days.
Cobalt does not contribute significantly to the spectra at 300 and 500 days,
and the strength of the cobalt lines decreases with time.
At 300 days only $\sim 7 \%$ of the radioactive cobalt remains.
Uncertainties in especially the cobalt 
collisional excitation and recombination data are therefore of minor 
importance.
The strongest cobalt line in the models is [Co~II] $1.019\mu$. 

While most line features in both the optical and IR are severely
blended, the IR lines are the ones least affected. For studies of line 
profiles these are therefore the most suitable. In particular the Ni II line
from stable $^{58}$Ni at $1.940 \mu$ is from this point of view 
especially interesting.

\begin{figure*}[t]
\includegraphics[width=140mm,clip,angle=-90]{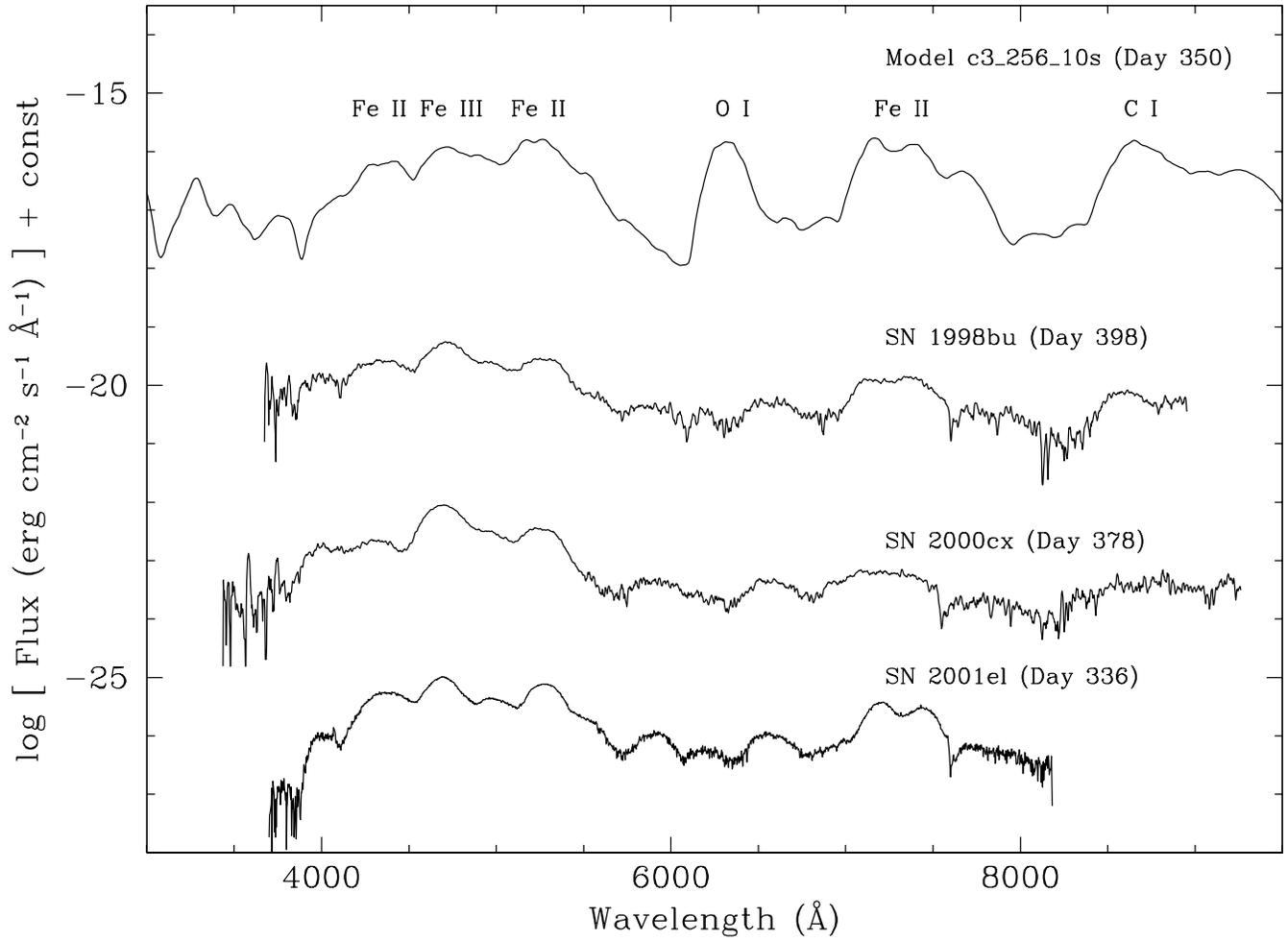}
\caption{Model spectrum at 350 days, for the c3\_3d\_256\_10s model compared 
to observations of SN 1998bu from day 398, SN 2000cx from day 378, 
and of SN 2001el from 336 days after explosion. 
For clarity the observed spectra have been shifted relative to each other,
the SN 1998bu spectrum by $-$4 dex, 
the SN 2000cx spectrum by $-$7 dex, and 
the SN 2001el spectrum by $-$10 dex. 
The observed spectra have been smoothed, dereddened, normalized to 350
days, and rescaled to a common distance of 10 Mpc as discussed in 
section~\ref{spectrum}.
}
\label{spec300d_00cx}
\end{figure*}

In Fig.~\ref{spec300d_00cx} we compare a model spectrum at 350 days
to observations of SN 1998bu on day 398, SN 2000cx on day 378, and SN 2001el
on day 336 after explosion. We have 
renormalized the flux of the three spectra to 350 days, using a decline 
rate of 0.0138 mags
per day in the V band (Sollerman et al. 2004). 
We have also scaled the observed
spectra to a common distance of 10 Mpc. 

The spectrum of SN 1998bu, taken from \cite{Spyro}, was obtained at
398 days after explosion. This is one of the best S/N spectra
available at this epoch, and one of the few covering the region above
8000 \AA. A problem is, however, that the spectrum is contaminated by
a light echo (Cappellaro \etal 2001), which especially in the blue gives a
substantial contribution to the flux. We have removed this as
discussed in \cite{Spyro}. For the
reddening of SN 1998bu we adopt $E_{\rm B-V}=0.30$ and a distance of 11
Mpc (Suntzeff \etal 1999). SN 1998bu has a normal luminosity and decline rate.

SN 2000cx was observed using VLT/FORS 360 days past maximum light.
These observations were performed with the 300V grism, an
order-sorting filter GG375 and a $1\farcs3$ wide slit.
The data were reduced in a standard way,
including flux-calibration using spectrophotmetric standard stars,
as outlined by Sollerman et al. (2004). 
For the smoothed spectrum of SN 2000cx shown in Fig.~\ref{spec300d_00cx} we
use a reddening $E_{\rm B-V}=0.08$ (Schlegel \etal 1998) and a distance of 
33 Mpc (Candia \etal 2003).
SN 2000cx was a rather peculiar SN Ia around maximum light, 
peculiarities including an unusual light-curve and
an unusual color evolution (Li et al. 2001).
However, as shown in Sollerman \etal (2004), at later
phases the light curve for this supernova actually behaved 
quite normally. In Fig.~\ref{spec300d_00cx} we also compare the 
late spectrum from SN 2000cx to the two other SN Ia and we
find that, at these epochs, the three spectra are quite similar.

The observations of SN 2001el were obtained with the same instrumental
set-up as for SN 2000cx at an epoch of 318 days past 
maximum.
For the reddening we adopt $E_{\rm B-V}=0.25$ and a
distance of 17.9 Mpc (Krisciunas et al. 2003).
SN 2001el was a rather normal SN Ia as far as its luminosity and
decline rate are concerned (Krisciunas et al. 2003), but it was the
first ``normal'' SN Ia that showed strong intrinsic polarization and a
high velocity detached Ca II IR triplet, interpreted as a clumpy shell
(Kasen et al. 2003).

The first thing to note from the comparison of the model with the
observed spectra is the general similarity of most of the strong
features in the spectrum. In particular, this applies to the Fe II-III
features, which are all seen, albeit with somewhat different relative
fluxes (see below). The absolute level of the model spectrum is lower than
the observed spectra,
which is explained by the low $^{56}$Ni mass in the c3\_3d\_256\_10s
model. 
Considering the limitations of the explosion models, we find the
general agreement of models and observations very promising.
With this background we now discuss the shortcomings of this
model.

Comparing the synthetic spectrum with the observed, it is obvious
that the dominant lines in the modeled spectrum, [O~I] $\wll~6300, 6364$ 
and [C~I]
$\wl~8727$, are not present in the observations.
This is particularly obvious for
the [O~I] line, where the peak is close to a 'valley' between two
emission features in the observed spectra. The [C~I] region is only
covered by the SN 1998bu and SN 2000cx spectra. 
There is in the SN 1998bu spectrum indeed a
feature close to the [C~I] line, but a closer inspection shows that
this is centered at $\sim 8600$ \AA, rather than at 8727 \AA. The
origin of this is therefore most likely Fe II $\wl 8621$, 
rather than [C~I]. The strong 
[C~I] and [O~I] lines are the most serious
discrepancies of the models, and show that these explosion models have
too large masses of these elements in the central regions of the
explosion model.  We discuss this further in section \ref{discussion}.

A less serious
discrepancy is that the c3\_3d\_256\_10s model is not
able to satisfactorily reproduce the characteristic three-bump feature
between 4500 \AA~  and 5500 \AA.  In particular, the observed strong bump
at $\sim$ 4700 \AA , which is mainly due to Fe~III 
(Fig.~\ref{specions513514}), is too weak in the
model, as can be seen in Fig.~\ref{spec300d_00cx}. This indicates
that the model has a too low degree of ionization.

Another slight discrepancy is the feature seen in the observations between 
5700 and 6200 \AA , but which is absent in the model. 
This region is in the model dominated by Fe~II emission. 
In addition Si~I, Fe~I, Fe~IV, Co~II, and Ni~II have transitions at these
wavelengths.

\begin{figure*}[t]
\includegraphics[width=140mm,clip,angle=-90]{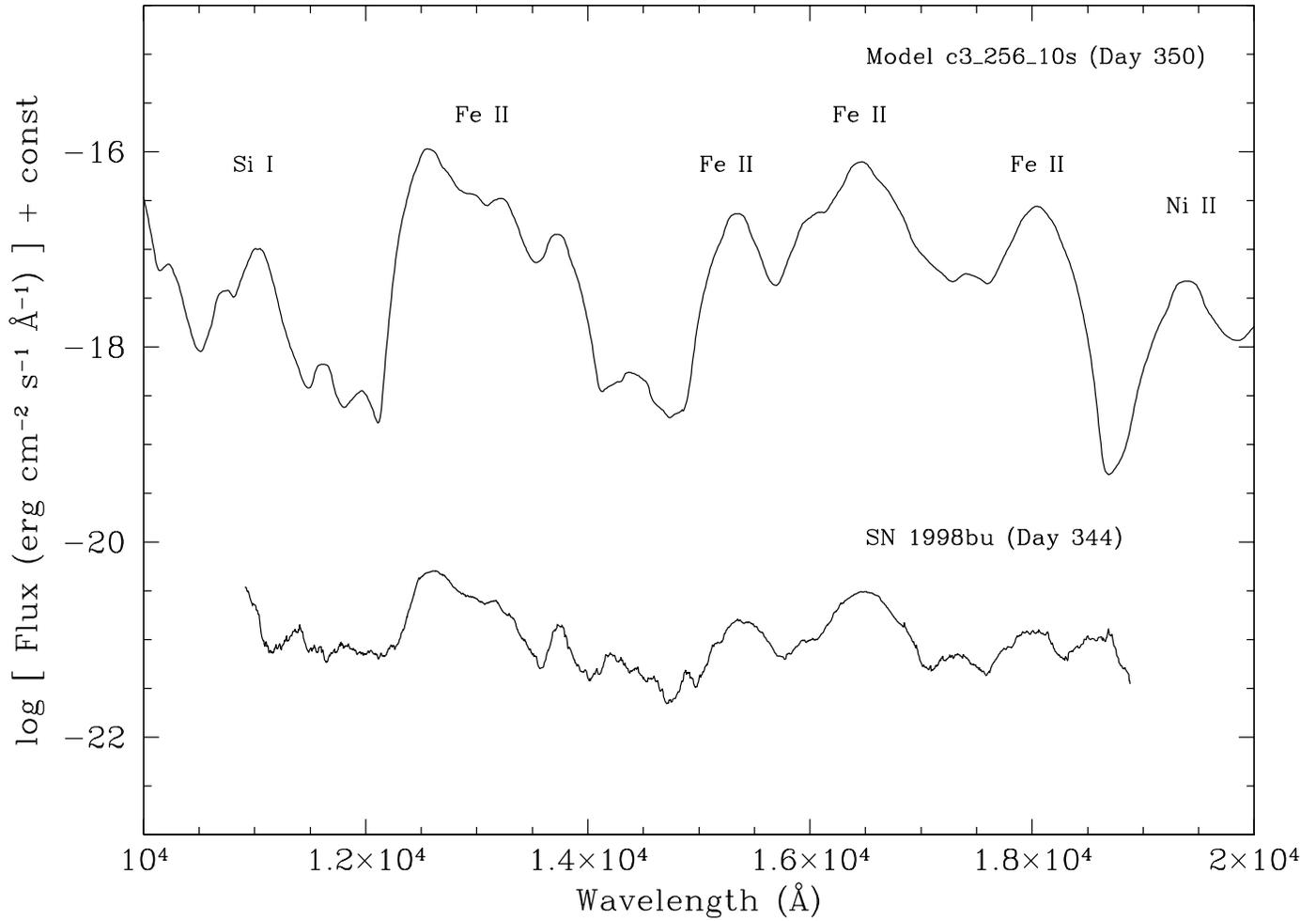}
\caption{Model IR spectrum at 350 days, for the c3\_3d\_256\_10s model compared 
to observations of SN 1998bu from day 344 (Spyromilio \etal 2004).
For clarity the observed spectrum has been shifted by $-$4 dex. 
The observations have been cleaned and heavily smoothed, using a
Savitzky-Golay polynomial smoothing filter
with a width of 50 wavelengthbins, corresponding to $\sim 4500 \kms$.
}
\label{spec98bu_ir}
\end{figure*}

In Fig.~\ref{spec98bu_ir} we compare the modeled IR spectrum at 350 days
to observations of SN 1998bu at 344 days from Spyromilio \etal (2004).
We find that the model in this wavelength region is 
able to reproduce most of the features seen in the observations. 
In particular, the relative strengths of the different Fe~II lines agree
very nicely with the observations. 
There is an indication that the [Si~I] $1.099 \mu$ line is also present. 
Unfortunately, the observed spectrum only covers half of the line.

The densities in the other three explosion models, which were only
calculated up to 1.2 -- 1.5 seconds, are significantly higher than for
the 10 second model, if scaled to the same epoch. If run to the
homologous stage they would, however, most likely end up with a lower
density. In particular, the b30\_3d\_768 model has the highest kinetic
energy and may therefore end up as the lowest density
model. Nevertheless, these models demonstrate that higher
densities result in lower temperatures and lower degrees of
ionization. Therefore, for these models the Fe III bump is completely
absent.

In the near-IR, the low degree of ionization in these models result in
a strong [Fe~I] $1.443 \mu$ line, which
is very weak in the c3\_3d\_256\_10s model. The same is true for the [Si~I]
$1.099\mu$ line. The fluxes of the O I and C I lines are in these
models considerably stronger. This demonstrates the importance of the
density structure of the explosion model to the late time spectra. The
fact that these models were not calculated to the homologous stage
therefore makes any conclusions on these models very uncertain.

To analyze the differences between the observations and model,
and because the spectrum reflects the temperature and ionization of the
ejecta, we now discuss in more detail how these quantities vary in the models.

\subsection{Temperatures}
\label{temps}

Because of the large range in density and composition of the ejecta, the
temperature and ionization will be highly inhomogeneous, depending on
the composition, 
density, and to a certain degree the geometrical position of the tracer 
particle.
This is one of the most important differences between 1D and 3D models.

At 300 days (500 days) the temperatures in the unburned C/O/Ne particles, 
vary from $\sim$2600 K to $\sim$8500~K (600 to 6000~K), with a 
characteristic temperature of $\sim$ 5000~K (3000~K).
The highest temperatures are found at the largest
radii. The reason is that the particles closest to the center, having 
the highest density, 
also undergo the fastest cooling (Axelrod 1980). Away from the center the densities
decrease, and the temperatures increase. 

\begin{figure*}[t]
\includegraphics[width=180mm,clip]{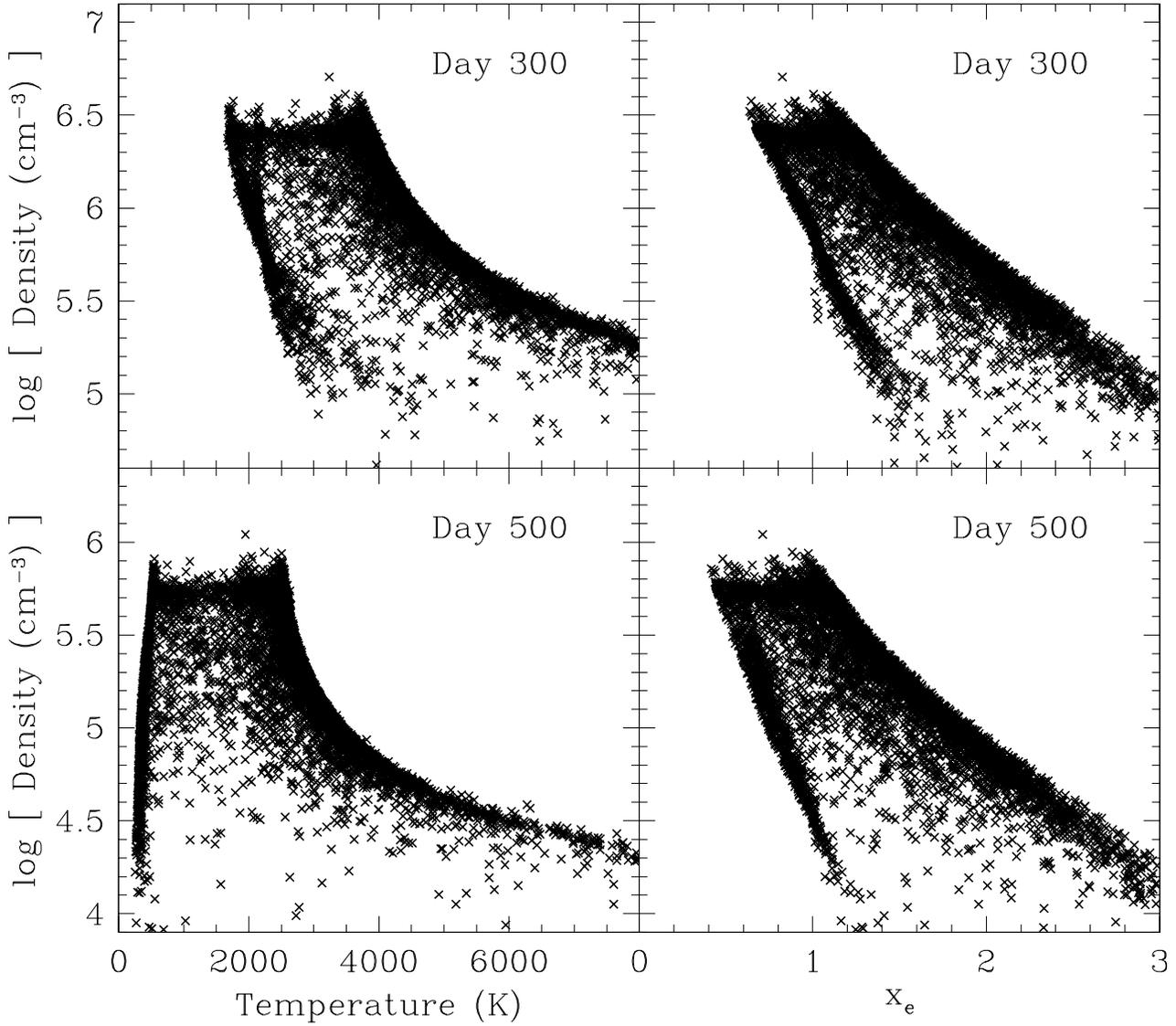}
\caption{
The crosses mark density and temperature (left hand panels) and 
density and electron fraction (right hand panels) for each of the Fe-rich 
mass elements 
the c3\_3d\_256\_10s model at 300 (upper panel) and 500 days (lower panel).
}
\label{texeden513514}
\end{figure*}

To further illustrate the highly inhomogeneous nature of the ejecta
conditions, we show in Fig.~\ref{texeden513514} the temperatures of 
the Fe-rich mass elements at 300 and 500 days, where we have plotted
the density and temperature for each of the Fe-rich mass elements. 
These particles have densities in the range
$\sim 3\EE{4}$ cm$^{-3}$ to $\sim 3\EE{6}$ cm$^{-3}$ at 300 days
(Fig.~\ref{dendist}).
At 300 days most of the mass
elements  are found in two temperature regions (along the two boundaries
seen in Fig.~\ref{texeden513514}),   
at $\sim$ 2000~K, and  $4000 - 5000$~K, respectively.

For a given 
density the variation in temperature, as seen in Fig.~\ref{texeden513514}, 
is due to different $^{56}$Ni-abundances
in the mass elements. The temperatures of the tracer particles 
show two boundaries in Fig.~\ref{texeden513514},
where most of the particles are found. 
These contain the most $^{56}$Ni-poor 
(lowest temperatures) and
most $^{56}$Ni-rich (highest temperatures) mass elements, respectively. 
For a given $^{56}$Ni fraction,
the temperature increases with decreasing densities.
The particles at the low-temperature boundary  
are composed of stable iron (mainly $^{54}$Fe) with a mass fraction 
of $\sim 0.7$, 
and stable nickel (mainly $^{58}$Ni) with a mass fraction of $\sim 0.3$. 
In these particles only 
$\sim 5\EE{-4}$ of the mass is radioactive $^{56}$Ni.  
At the high-temperature boundary, on the other hand, the mass fraction of
$^{56}$Ni is $\sim 0.9$. 

The nucleosynthesis taking place within a tracer particle is determined
mainly by the peak temperature and density of the particle. If the temperature
is sufficiently high,  T$\gsim 6\EE{9}$~K (Travaglio \etal 2004), 
nuclear statistical equilibrium (NSE) is reached, resulting in a composition 
dominated by iron-peak elements.
The final composition of a particle reaching NSE depends on   
both on the temperature and density and also on 
the neutron excess ($\eta$),
 which is detemined by the amount of electron capture taking place.
The higher the neutron excess the more neutron rich isotopes such as
$^{58}$Ni and $^{54}$Fe form.

At 500 days the temperatures have decreased 
significantly, as seen in Fig.~\ref{texeden513514}.
The temperatures are now peaked 
at $\sim$ 500~K and $\sim$ 3000~K. 
For the $^{56}$Ni-rich particles 
the density and temperature still correlate, so that
the temperatures in the mass elements with lower densities are 
higher than in the mass elements of higher densities.

For the $^{56}$Ni-poor particles on the other hand, 
the temperatures of the particles are more or less constant, 
independent of density.
The reason is that at these low temperatures 
cooling by far-IR Fe~II fine structure lines is becoming important. 
For the high density, low temperature regions, the ionization
is lower (see the right panels in Fig.~\ref{texeden513514}), 
resulting in a higher 
abundance of Fe~I than of Fe~II, and a less 
efficient cooling. 
Therefore, for these particles the cooling is
determined both by the ratio of Fe~I/Fe~II and density. 

The spread in temperatures for the Fe-rich particles 
in the b30\_3d\_768, b5\_3d\_256, 
and the c3\_3d\_256 models are smaller than for the c3\_3d\_256\_10s model. 
This depends mainly on the differences
in densities. The higher the density is, the smaller the
spread in temperature will be.

Also for particles not burned all the way to NSE, rich in 
intermediate mass elements, we find that particles with similar compositions
gather along certain boundaries. The temperatures for some of these
particles are even higher than for the Fe-rich particles rich in
$^{56}$Ni, due to the less efficient cooling in these particles.

\subsection{Ionization}
\label{ions}

In the right panels of Fig.~\ref{texeden513514} we show
the electron fraction 
as function of density within each Fe-rich mass element
at 300 and 500 days for the c3\_3d\_256\_10s model.
The electron fraction, x$_e$, is here defined
as the ratio of 
the number density of free electrons, n$_e$,
to the total number density of neutral atoms and ions, n,
i.e. x$_e$ = n$_e$/n.
At 500 days the electron fraction varies from $\sim$ 0.4 to $\gsim$ 
3, depending on density.
The lower boundary of the electron fractions at 300 days is a factor
of $\sim 1.5$ higher. 
The same boundaries as seen for the temperature are present 
for the electron fraction in Fig.~\ref{texeden513514}.
This means that the upper boundaries in the plots include the 
most $^{56}$Ni-rich 
particles, while the mass elements along the lower boundary
contains the least $^{56}$Ni. As is clearly seen in these figures, 
a decrease in density results in an increase in electron fraction
(for a given $^{56}$Ni mass) due to reduced recombination rates.

As expected, the degree of ionization decreases as the energy input decreases.
At both epochs Fe~II is the dominant ion. However, at 500 days
the Fe~I abundance is $\gsim\ 0.1$, in the high density regions.
Conversely, Fe~III becomes increasingly important as energy input 
increases.

Similar to the temperature, the spread in electron fraction for the 
Fe-rich particles in the b30\_3d\_768, b5\_3d\_256, 
and the c3\_3d\_256 models
are smaller than for the c3\_3d\_256\_10s model. 
The higher densities in these models result in a lower degree of
ionization.

For all models we have assumed that the positrons deposit their energy 
locally in the mass elements
containing the newly synthesized nickel. 
If we instead  assume non-local
deposition of the positrons (Milne \etal 1999, 2001), 
the Fe~III abundance would decrease, making 
the ionization problem even worse. On the
other hand, the degree of ionization in the unburned regions would increase
somewhat, reducing the strength of the O I and C I lines. This effect
is, however, probably small.

\section{Discussion}
\label{discussion}

Models of early Type Ia supernova spectra, based on 3D 
hydrodynamical models, have been developed 
by Thomas \etal (2002), and
Thomas (2003). They have used a modified version of the spectrum code 
SYNOW to include effects of non-spherical symmetric geometries.
In particular, Thomas \etal have studied the effects of non-spherical 
inhomogeneities on the line profiles. This has in contrast to this
paper mainly been in the context of early stages, and at high velocities.
In Thomas \etal (2004) the high velocity Ca~II features seen 
in SN 2000cx near maximum light are modeled, based on 3D ejecta models.
They find that 3D models are required to explain the Ca~II features, 
and they also estimate the mass of the high velocity ejecta.

Our conclusions with regard to unburned carbon and oxygen in the ejecta are
complementary to previous studies using early epoch spectra.
Observations in the near infrared region 
(Marion \etal 2003) 
of twelve supernovae near maximum light, 
show no sign of unburned carbon at velocities above $\sim 3000 \kms $.
Based on these observations they put a limit to the amount of unburned matter 
to $\lsim$ 10\% of the progenitor mass.

By comparing observed spectra of SN 2002bo to 1D synthetic
spectra at different times, both around maximum light and
in the nebular phase, Stehle \etal (2004) derive the abundance distribution 
of the supernova ejecta. They find a $^{56}$Ni mass of 0.52 \msun\ and 
evidence for intermediate mass elements at high velocities.  
They find no sign of carbon lines at any velocities, and inferred 
an upper limit of the total mass of carbon to $1.6\EE{-3}$~\msun . The 
corresponding mass for oxygen was 0.11 \msun\ where it is most dominant
at high velocities, and no oxygen was observed below $6000 \kms$.

Further, in order to examine the mixing of unburned material to low
velocities, as seen in 3D deflagration models, 
Baron \etal (2003) 
modified the composition of the 1D, 
deflagration model W7 
(Nomoto \etal 1984; Thielemann \etal 1986) by introducing unburned
material into the central regions. They modeled the spectra from 3000
-- 8000 \AA\ at 15-25 days past maximum, and concluded that it was, 
from their models, not possible to rule out the presence of unburned
material at low velocities.  As a consequence of this, they pointed
out the importance of modeling of late time nebular spectra, based on
actual 3D hydrodynamical models, in order to resolve these questions.
This is demonstrated by our calculations.

\begin{figure*}[t]
\includegraphics[width=150mm,clip]{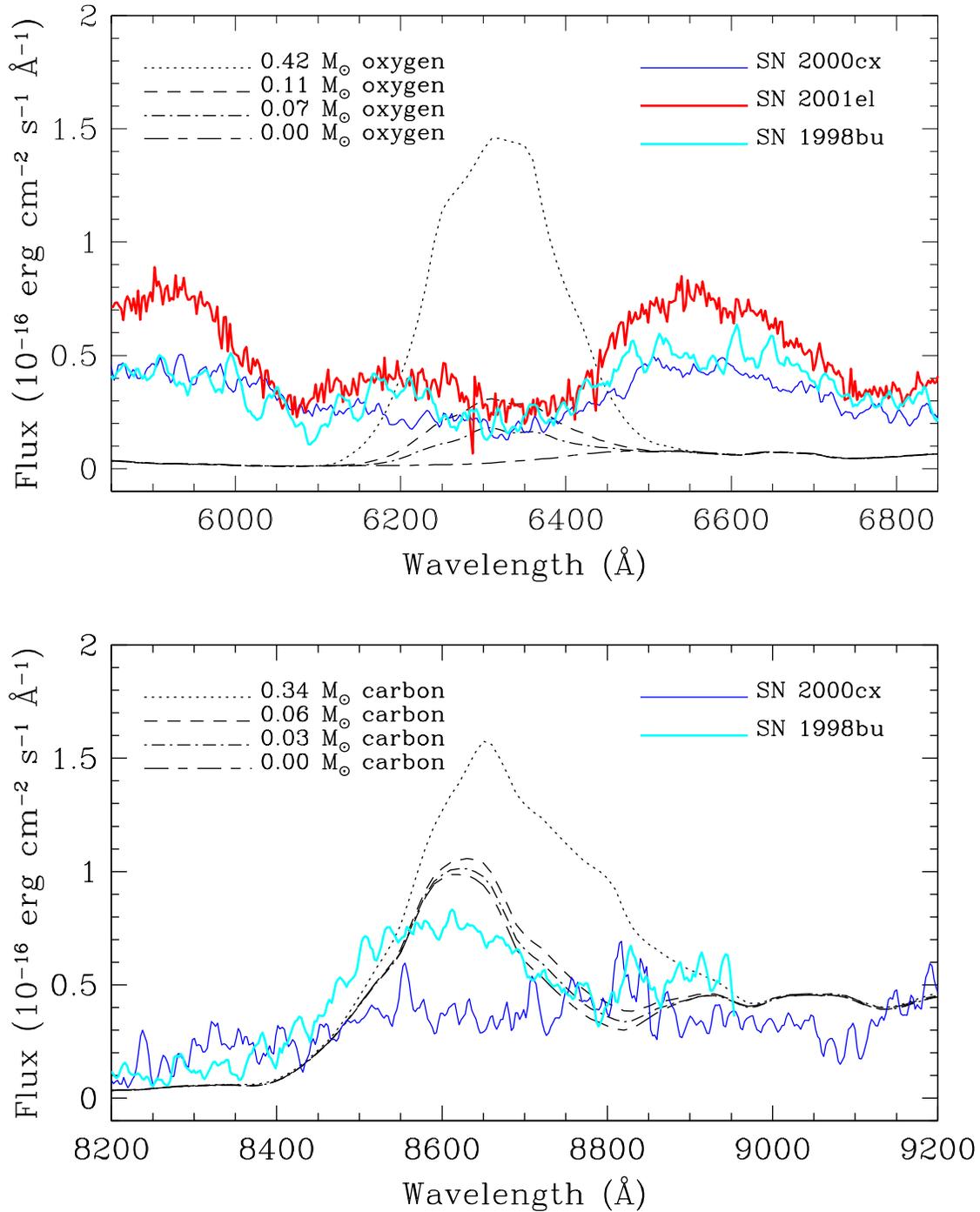}
\caption{
The region around the [O~I] $\wll~6300, 6364$ and [C~I] $\wl~8727$ lines
based on the c3\_3d\_256\_10s model at 350 days together with observed spectra 
for SN 2000cx, SN 2001el and SN 1998bu.
The original model, containing 0.42~\msun\ of oxygen and 0.34~\msun\ 
of carbon (dotted curves), 
is compared to models where the masses of oxygen and carbon
have  been artificially reduced. 
The dot-dashed curve includes only the partially burned regions, with
$\sim0.03$~\msun\ carbon and $\sim0.07$~\msun\ oxygen, while the dashed curve also has
additional 0.03~\msun\ carbon and 0.04~\msun\ oxygen from unburned material.
The remaining feature in the oxygen and carbon free model 
at $\sim 8600$ \AA\  is due to Fe~II. 
}
\label{specub}
\end{figure*}

The synthetic nebular spectra presented in this paper show strong
carbon and oxygen lines in contradiction to our observations. 
This fact
demonstrates that the amount of unburned material in the center of these explosion
models is too large.  In order to estimate an upper limit to the
mass of unburned matter in the central region allowed from the nebular
spectra, we have recalculated the spectra from the c3\_3d\_256\_10s model
with varying amounts of unburned matter. The results are shown in
Fig.~\ref{specub}, where we show the spectra in the wavelength regions
containing the [O~I] $\wll~6300, 6364$ and [C I] $\wl 8727$ lines,
together with observations of SN 1998bu, SN 2000cx, and SN 2001el.
The absolute flux calibration of the spectra of SNe 1998bu and 2000cx is
performed by comparison to contemporary broad band photometry. For SN
2001el we use a late time magnitude based on the assumption that
this supernova has the same time-evolution  ($\Delta$~m$^{\rm 350}_{\rm V}$)
as the well observed SN 1996X. We integrate the spectra under the V band
filter function and scale it to match the extrapolated late V band
magnitude.
Finally, the spectra are corrected for Galactic extinction, transferred
to a common distance and scaled to a common epoch.

The dotted curve in Fig.~\ref{specub} shows the original model. The model 
contains in total 0.42 \msun\ of oxygen and 0.34 \msun\ of carbon, 
residing both in the unburned particles and in the partially 
burned particles.
Note that the observed spectra are calibrated on an absolute scale. 
We can therefore directly compare the fluxes in the O~I and C~I lines
with the observations, and not only relative to the 'continuum'.

As an extreme we have artificially removed all carbon and oxygen 
emission in one model, shown as the lowest short-long dashed line
in Fig.~\ref{specub}.
We also show two models where we have varied
the number of unburned particles. In the first  
we removed all unburned particles, including only the carbon
and oxygen residing in the partially burned particles 
(0.07 \msun\ of oxygen and 0.03 \msun\ of carbon).
In the second we reduced the number of unburned particles by a factor of 
10, resulting in oxygen and carbon masses of 0.11 \msun\ and 0.06 \msun, 
respectively.
From this figure it is seen that the original model is in clear 
disagreement with the observations for both the C~I and O~I lines.
The model with only the carbon and oxygen in the partially burned
particles is just compatible with the observations. As an upper limit to the 
{\it unburned} mass we estimate this to be close to the 0.03 \msun\ of carbon
and 0.04 \msun\ of oxygen in the dashed line model in Fig.~\ref{specub}. This
is in addition to the carbon and oxygen in the partially burned gas. We
do stress that these limits only apply to the low velocity gas in the core.

Although we do not find any signs of oxygen in our observations there
are early claims in the literature of detection of [O~I] $\wll~6300, 6364$  
in SN 1937C (Minkowski, 1939). This has, however, not been confirmed by modern 
observations.

\begin{figure*}[t]
\includegraphics[width=170mm,clip]{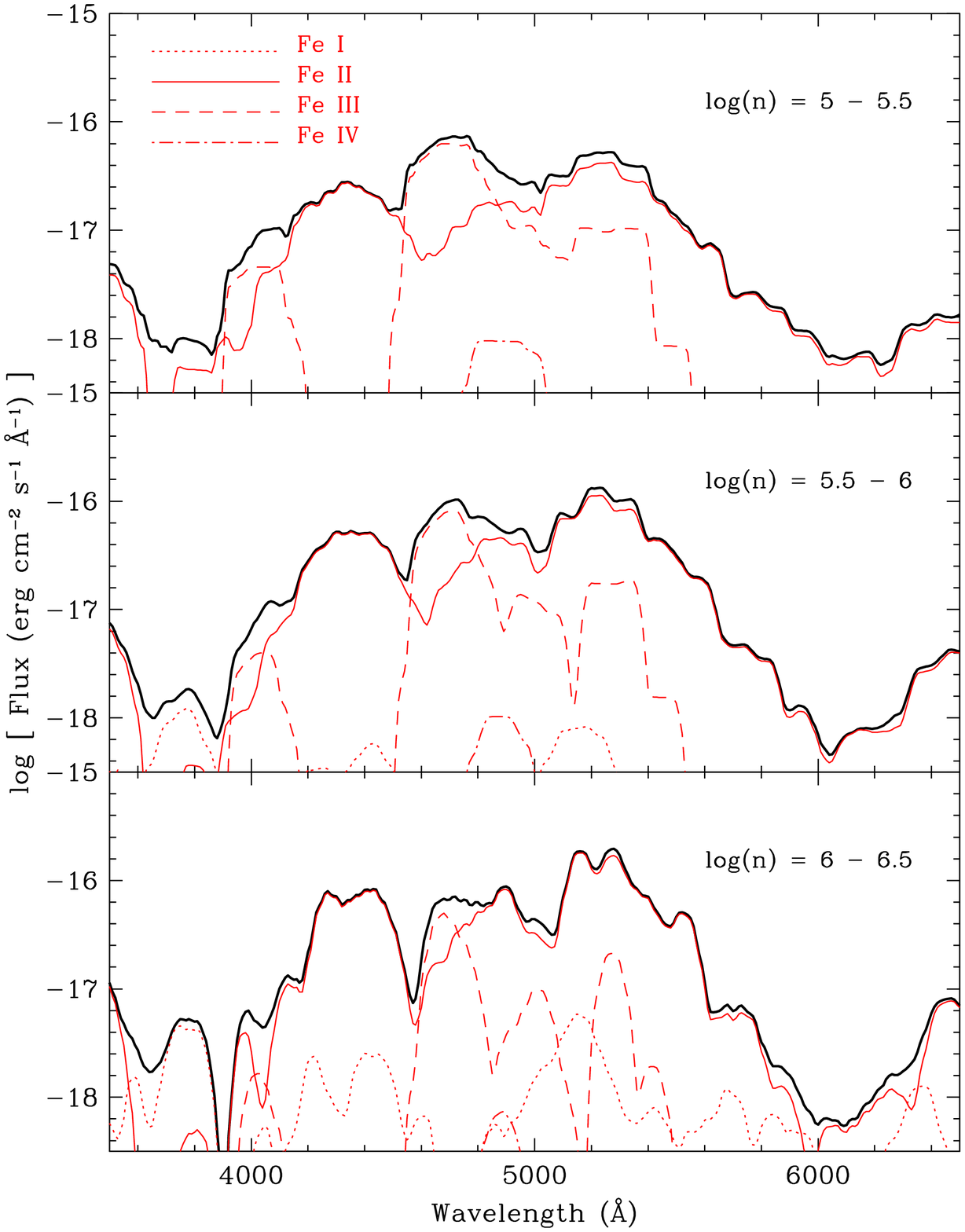}
\caption{
Spectra based on the c3\_3d\_256\_10s model at 300 days. 
The upper panel shows the spectrum from the 
Fe-rich particles
in the density range $10^5 - 3\times10^5$ cm$^{-3}$, the middle spectrum
from those in the range
$3\times10^5 - 10^6$ cm$^{-3}$ and the lower panel the spectrum from
those in the range $10^6 - 3\times10^6$ cm$^{-3}$. The thick, solid
line gives the total flux from these density ranges. 
}
\label{specden}
\end{figure*}

In the wavelength range 4500 -- 5500 \AA\ the observations show three
characteristic bumps (Fig.~\ref{spec300d_00cx}). These can all be seen
in the model, and are explained as a mixture of Fe~II and Fe~III
emission.  The relative strengths of the Fe~II and Fe~III features are
sensitive to the densities in the emitting gas.  In Fig.~\ref{specden}
we show the contributions from the different ionization stages of iron
to this wavelength region for particles of different density
ranges. In the upper panel we show the contribution from the Fe-rich
tracer particles in the range $(1-3)\times 10^5$~cm$^{-3}$, and in
the two lower panels for the density ranges $(3-10)\times10^{5}$~cm$^{-3}$ 
and $(1-3)\times 10^6$~cm$^{-3}$.  The
contributions to the total flux in this wavelength region from the
above density ranges are 17 \%, 34 \% and 49 \%, respectively.

Fig.~\ref{specden} clearly shows the effects of varying the density.  Especially
the strength of the Fe~III feature around 4600 -- 4800 \AA\ is sensitive to the
density of the emitting gas.  An increase in density reduces the degree
of ionization, due to increased recombination rates, and consequently the
Fe~II abundance increases relative to Fe~III. 
In the c3\_3d\_256\_10s model both the Fe~II peaks at $\sim$~4300~\AA\ and at 
$\sim$~5200~\AA\ , as well as the Fe~III peak, are clearly seen in 
Fig.~\ref{spec300d_00cx}. Compared to the observed spectra, the relative
fluxes of the two Fe~II peaks are well reproduced. The ratio of the flux in 
the Fe~II and Fe~III peaks is, however, a factor of $\sim 2-3$ 
too large compared to the observations. A decrease in the density compared
to the c3\_3d\_256\_10s model is therefore indicated from these observations.
Also the too strong Fe~II complex at 7000 -- 8000 \AA, and the blend of 
Fe~II and C~I at 8500 -- 9500 \AA\,  
indicate a too low ionization. In addition, the large dip in the model
spectrum at $\sim$ 6000 \AA\ would be reduced by a lower density.

In the other three non-homologous models 
the densities are even higher, and Fe~II is totally dominating the spectrum in
this wavelength region. From this we see the importance of following the
hydrodynamical calculations until homologous expansion is reached.

In addition to decreasing the densities, an alternative way to increase the
ionization is to increase the amount of $^{56}$Ni formed in the explosion.
The amount of $^{56}$Ni in the c3\_3d\_256\_10s model is only 0.28 \msun.  The
reason for this low value (which is not typical for pure deflagration models)
was discussed in Sec. 2.1.  In Table \ref{modtab} the masses of $^{56}$Ni are
given for all the models on which the present paper is based. In the model
with the highest resolution (model b30\_3d\_768) also the largest mass of
$^{56}$Ni is formed, i.e. 0.42 \msun.  These values are still at the low end 
of what
is derived from observations. Contardo \etal(2000) estimate $^{56}$Ni-masses
from observed UBVRI light curves for a number of type Ia supernovae, and find
M($^{56}$Ni) = 0.40 -- 1.1 \msun, except for SN 1991bg with M($^{56}$Ni) = 0.11
\msun.  For comparison, the popular W7 model has M($^{56}$Ni) = 0.60 \msun.  A
higher $^{56}$Ni abundance increases the ionization, and thus would improve the
agreement with the observed spectra shown in Fig.~\ref{spec300d_00cx}.

To test this hypothesis we have made one-dimensional spectrum
calculations based on the W7 model, which will be discussed in more
detail in \cite{KF05}. We find that at 300 days this model reproduces
the observations quite well. In particular, the [Fe~III] features are
now better reproduced.
At 300 days the densities in the Fe-rich zones in the W7 model vary
from $1.2\EE5$ cm$^{-3}$ to $1.5\EE6$ cm$^{-3}$.  This range
corresponds to the upper two spectra in Fig. \ref{specden}.  From this
we conclude that the reason why the W7 model reproduces the observed
spectra better than the c3\_3d\_256\_10s model is the
combination of a lower mean density and a higher $^{56}$Ni mass.

If we compare the synthetic spectra based on the
non-homologous c3\_3d\_256 and b30\_3d\_768 models at 300 and 500 days
(not shown here), we find that both models show too strong O~I and C~I
lines.  The $^{12}$C and $^{16}$O masses for the different models are
given in Table~\ref{modtab}.  While there is a considerable difference
in these masses depending on the resolution and initiation of the
burning, this is not enough to solve the problem with the strong O I
and C I features. This could be due to the fact that model
b30\_3d\_768 was not evolved far enough in time. But it could also
reflect a deeper problem of the models.

There are several assumptions and uncertainties in the 3D hydro calculations
that might affect the still existing differences between the computed late
time spectra and our observations. First of all, low-resolution models such as the
c3\_3d\_256\_10s model have been computed with the aim to analyze differences in
the model predictions resulting from different physical parameters, for
instance the C-to-O ratio (R\"{o}pke \& Hillebrandt 2004), the metallicity, and the ignition density.  They
are not meant to give realistic models of the explosion.  Models with higher
resolution (such as model b30\_3d\_768) are more realistic, and for given
initial conditions they seem to be almost ``converged'' as far as global
properties are concerned, i.e. the Ni-mass and the explosion energy. But these
models have not yet been evolved far enough to make safe predictions for the
density, the temperature, and the element distribution in velocity
space. However, we can predict from model b30\_3d\_768 that in models with
higher numerical resolution the amount of unburned material at low velocities
will be less and more $^{56}$Ni will be formed. 

Additional uncertainties result from the (unknown) ignition conditions, the 
still rather ad-hoc sub-grid scale model for the burning velocity (Schmidt 
\etal 2005), the poorly known physics of thermonuclear burning in the
distributed regime at low densities (R\"{o}pke \& Hillebrandt 2005a), or the
need to perform full-star computations in order to avoid artificial
symmetries (R\"opke \& Hillebrandt 2005b). 
Whether or not all these necessary modifications of the models
will reduce the amount of unburned carbon and oxygen to the low 
values indicated
by the observed nebular spectra is an open question, and will be addressed
in future papers.

Another way to avoid the low velocity carbon and oxygen, as well as increasing
the $^{56}$Ni-mass, could be to invoke a delayed detonation (Khokhlov 1991;
Yamaoka \etal 1992; Niemeyer \& Woosley 1997).
Light curve modeling by H\"{o}flich \& Khokhlov (1996),
based on 1D simulations, has shown that delayed detonations give good fits to
the light curves and early spectra, provided the transition density from the
deflagration to the detonation phase is properly adjusted, introducing a new
free parameter. Recently Gamezo \etal (2004a, 2004b) found in a 3D model that a
transition from deflagration to detonation can remove efficiently low velocity
carbon and oxygen. In addition, the mass of $^{56}$Ni increased considerably.

Plewa \etal (2004) present a gravitationally confined detonation
mechanism in which a single rising deflagration bubble 
triggers a detonation near the stellar surface.  
These models are thought to result 
in mildely asymmetric explosions, and with 
 energetics and chemical composition
similar to those of the delayed detonation models.

It is, however, in our view premature to draw any strong conclusions
of a preferred explosion mechanism in either direction. In particular,
it is not at all clear what kind of physics might cause a transition
to a detonation. On the contrary, all investigations carried out so
far seem to rule out the mechanisms responsible for
deflagration-to-detonation transitions in laboratory combustion for
SNe Ia (Niemeyer 1999; Lisewski \etal 2000; R\"opke \etal 2004a, 2004b; Bell
\etal 2004). Moreover, it is unclear even whether or not a detonation
ignited in one pocket of unburned C-O fuel can incinerate the rest of
the already fast expanding star. Because the detonation front in a SN
Ia is weak and propagates with sonic velocity pressure waves may not
be able to penetrate through regions consisting of burned material. It
is quite possible that numerical diffusion has caused the almost
complete burning found in the simulation of Gamezo \etal (2004a, 2004b)
(Niemeyer, Livne, private communication).

\section{Summary}
\label{summary}

Our main aim in this paper is to show the potential of testing
multidimensional explosion models by detailed modeling of the emission
later than $\sim$ 100 days after the explosion. To do this we have
calculated late time spectra based on 3D hydrodynamical and
nucleosynthesis simulations.  We have mainly studied one 
model which has been calculated to the homologous stage, c3\_3d\_256\_10s,
at two different epochs, 300 and 500 days, but also investigated three
other less evolved models, to get an idea of how the still limited
numerical resolution of today's 3D models and the initiation of
burning might affect the conclusions.

Our results can be summarized in the following points:

\begin{itemize}

\item 
The observed features in both the optical and IR spectra, dominated 
by Fe~II-III emission, are
qualitatively well reproduced, given the minimum of free parameters in
the explosion models. \\

\item 
The models, however, as they stand predict strong carbon and oxygen
lines, in particular [C~I] $\wl~8727$, [C~I] $\wll~9824, 9850$ and
[O~I] $\wll~6300, 6364$, at both epochs.  
The strong [O~I] in the model predictions is clearly in conflict with the 
observations. As far
as [C~I] is concerned the conclusions are based only on a few spectra.
Unfortunately there exist only a few published SNe Ia late spectra
which extend to the red beyond 8000 \AA .  Nonetheless, from the
comparison between the models and observations we think we can set
an upper limit of the {\it unburned} mass inside of $\sim 10^4~\kms$ of
$\sim$ 0.07 \msun\ for the published cases. We emphasize that this only
applies to material lower than this velocity. At higher velocities late
epoch spectra are less sensitive because of the low gamma-ray deposition
rate there. \\ 

\item 
In the models the temperature and ionization in the different mass
elements vary strongly, depending mainly on the composition and density of the
mass elements.  From 300 days to 500 days the temperatures decrease
significantly. At 500 days a fraction of the Fe-rich mass elements starts to
undergo the IR-catastrophe, and their temperatures have dropped to 
$\sim 500$~K. At the same epoch there are regions with temperatures $\gsim 5000$~K.
This illustrates the need of detailed modeling based on 3D explosions, 
and shows that 1D models for the spectra are a poor approximation for
pure deflagration models. \\

\item 
The ionization in the models used in this study is somewhat too low in
comparison to the supernovae for which late spectra exist. This can be
seen from the Fe~II-III features in the wavelength interval 4500 -- 5500 \AA .
By increasing the mass of $^{56}$Ni and/or reducing the densities, we would
get a higher ionization and a better agreement with observations.  Model
b30\_3d\_768, if evolved into homologous expansion, would give us lower
densities as requested, indicating the need for such elaborate (and expensive)
models. Again, the Fe~II-III features in the late nebular spectra provide
excellent measures to test the explosion models based on the observations of
individual supernovae. \\

\end{itemize}

To conclude, we have demonstrated that modeling of late spectra can put strong
constraints on the hydrodynamics and nucleosynthesis in the explosion.  In the
future we intend to explore more realistic explosion models, as well as a more
accurate scheme for radiative transfer in the optical range.  The calculations
presented here should in this context be seen as a first step. \\

\noindent
{\em Acknowledgements.}
We are grateful to Ken Nomoto for helpful discussions, to Sultana
Nahar for supplying atomic data in advance of publication, and to 
David Branch for valuable comments and suggestions. 
Financial support for this work was
provided to the Stockholm group by the Swedish National Space Board and Swedish
Research Council. 
This work was supported in part by the European Research Training 
Network "The Physics of type Ia Supernova Explosions'' under contract 
HPRN-CT-2002-00303.

\clearpage

\begin{table*} [t]
\begin{tabular}[t]{|l||l|l|l|l|l|l|l|}
\hline
Model & Mode of ignition & Grid size & Time & M($^{56}$Ni) & M(unburned O) & M(unburned C) & Kinetic energy \\
&&& (s) & (\msun) & (\msun) & (\msun) & (ergs) \\
\hline
b30\_3d\_768      & 30 bubbles &  $768^3$ & 1.2 & 0.42 & 0.26 & 0.28 & $6.7\EE{50}$\\
b5\_3d\_256       & 5 bubbles  &  $256^3$ & 1.5 & 0.33 & 0.29 & 0.24 & $5.1\EE{50}$\\
c3\_3d\_256       & central    &  $256^3$ & 1.5 & 0.31 & 0.34 & 0.35 & $4.4\EE{50}$\\
c3\_3d\_256\_10s      & central    &  $256^3$ & 10  & 0.28 & 0.35 & 0.31 & $\sim 4.4\EE{50}$\\
\hline
\end{tabular}
\\
\caption{The explosion models studied in this paper from 
Travaglio \etal (2004),  and from
R\"{o}pke (2005). The time in this table 
is the time for which the hydrodynamics and nucleosynthesis
has been calculated. 
}
\label{modtab}
\end{table*}

\clearpage

\clearpage

\clearpage

\clearpage

\begin{thebibliography}{}
\bibitem{A80} Axelrod, T.~S. 1980, Ph.D. thesis, Univ. California, Santa Cruz
\bibitem{BLH03} Baron, E., Lentz, E.~J., \& Hauschildt, P.~H. 2003, ApJ, 588, L29
\bibitem[Bell et al.(2004)]{bell}  Bell,~J.~B., Day M.~S., Rendleman, C.~A., Woosley, S.~E., \& Zingale M. 2004, ApJ, 608, 883
\bibitem[Branch et al.(2005)]{branch}  Branch,~D., Baron,~E., Hall,~N., 
Melakayil,~M., \& Parrent,~J. 2005, PASP in press, astro-ph/0503165
\bibitem[Candia et al.(2003)]{Cand} Candia, P., Krisciunas, K., Suntzeff, N. B., \etal  2003, PASP, 115, 277
\bibitem[Cappellaro et al.(2001)]{Cap} Cappellaro, E., et 
al.\ 2001, \apjl, 549, L215 
\bibitem{CLV00} Contardo, G., Leibundgut, B., \& Vacca, W.D. 2000, A\&A, 359, 876 
\bibitem{G03} Gamezo,~V.~N., Khokhlov,~A.~M., Oran,~E.~S.,
  Ctchelkanova,~A.~Y., \& Rosenberg,~R.~O. 2003, Science, 299, 77 
\bibitem[Gamezo, Khokhlov, \& Oran(2004)]{GKO04a} Gamezo, 
V.~N., Khokhlov, A.~M., \& Oran, E.~S.\ 2004a, Physical Review Letters, 92, 
21
\bibitem[Gamezo, Khokhlov, \& Oran(2004)]{GKO04b} Gamezo, 
V.~N., Khokhlov, A.~M., \& Oran, E.~S.\ 2004b, submitted to ApJ, 
astro-ph/0409598 
\bibitem[Garcia-Senz and Bravo(2005)]{GSB05}  Garcia-Senz,~D., \& 
Bravo,~E.  2005, A\&A 430, 585
\bibitem{GSW95} Garcia-Senz,~D., \& Woosley,~S.~E.\ 1995, ApJ 454, 895 
\bibitem[Hoeflich \& Khokhlov(1996)]{1996ApJ...457..500H} H\"{o}flich, P.~\& 
Khokhlov, A.\ 1996, \apj, 457, 500 
\bibitem[H{\" o}flich \& Stein(2002)]{2002ApJ...568..779H} H{\" o}flich, 
P.~\& Stein, J.\ 2002, \apj, 568, 779 
\bibitem{IBN99} Iwamoto, K., Brachwitz, F., Nomoto, K. I., et al. 1999, ApJS, 125, 439
\bibitem[Khokhlov(1991)]{1991A&A...245..114K} Khokhlov, A.~M.\ 1991, \aap, 
245, 114 
\bibitem{Ketal03} Kasen, D., Nugent, P., Wang, L., Howell, D.~A., Wheeler, J.~C., 
H\"{o}flich, P., Baade, D., Baron, E.,  \& Hauschildt, P.~H. 2003, ApJ 593, 788
\bibitem{KF92} Kozma, C. \& Fransson, C. 1992, ApJ, 390, 602 
\bibitem{KF98a} Kozma, C. \& Fransson, C. 1998a, ApJ, 496, 946 
\bibitem{KF98b} Kozma, C. \& Fransson, C. 1998b, ApJ, 497, 431 
\bibitem[Kozma \& Fransson (2005)]{KF05} Kozma, C. \& Fransson, C. 2005, in preparation 
\bibitem{KSC03} Krisciunas, K., Suntzeff, N.~B., Candida, P., \etal 2003, AJ, 125, 166
\bibitem{K88} Kurucz, R.~L. 1988, Trans. IAU, XXB, ed. M. McNally,
  Dordrecht: Kluwer, p. 168. 
\bibitem{LM00} Langanke, K., \& Martinez-Pinedo, G. 2000, Nucl. Phys. A, 673, 481
\bibitem{Liea01} Li, W., Filippenko, A.~V., Gates, E., \etal 2001, PASP, 113, 1178
\bibitem[Lisewski, Hillebrandt, \& Woosley(2000)]{2000ApJ...538..831L} 
Lisewski, A.~M., Hillebrandt, W., \& Woosley, S.~E.\ 2000, \apj,  538, 831
\bibitem{LJS98} Liu, W., Jeffery, D.~J., \& Schultz D.~R. 1998, ApJ, 494, 812 
\bibitem{MHVW03} Marion, G.~H, H\"{o}flich, P., Vacca, W.~D., \& Wheeler, J.~C. 2003, ApJ, 591, 316
\bibitem{MLD00} Martinez-Pinedo, G., Langanke, K., \& Dean, D. J. 2000, ApJS, 126, 493
\bibitem{MTL99} Milne, P.~A., The, L.-S., Leising, M.~D. 1999, ApJS,  124, 503
\bibitem{MTL01} Milne, P.~A., The, L.-S., Leising, M.~D. 2001, ApJ, 559, 1019
\bibitem{M39} Minkowski,~R. 1939, ApJ, 89, 156 
\bibitem{N96} Nahar, S.~N. 1996, Phys. Rev. A, 53, 2417 
\bibitem{N97} Nahar, S.~N. 1997, Phys. Rev. A, 55, 1980 
\bibitem{NBP97} Nahar, S.~N., Bautista, M.~A., \& Pradhan, A.~K. 1997,
  ApJ, 479, 497 
\bibitem[Niemeyer(1999)]{1999ApJ...523L..57N} Niemeyer, J.~C.\ 1999, \apjl, 
523, L57 
\bibitem{NW97} Niemeyer, J.~C., \& Woosley, S.~E. 1997, ApJ, 475, 740
\bibitem{NTY84} Nomoto, K., Thielemann, F.-K., \& Yokoi, K. 1984, ApJ, 286, 644 
\bibitem{PRUJ98} Pickering, J.~C., Raassen, A.~J.~J., Uylings, P.~H.~M., \& Johansson, S. 1998, APJS, 117, 261
\bibitem{Q98} Quinet, P. 1998, A\&AS, 129, 147 
\bibitem{RHN02a} Reinecke, M., Hillebrandt, W., \& Niemeyer, J.C. 2002a, A\&A, 386, 936 
\bibitem{RHN02b} Reinecke, M., Hillebrandt, W., \& Niemeyer, J.C. 2002b, A\&A, 391, 1167 
\bibitem{PCL04} Plewa,~T., Calder,~A.~C., \& Lamb,~D.~Q. 2004, ApJ, 612, 37
\bibitem{R04} R\"{o}pke,~F.~K. 2005, A\&A, 432, 969
\bibitem{RH04} R\"{o}pke,~F.~K., \& Hillebrandt, W. 2004, A\&A, 420, L1
\bibitem{Rea041} R\"{o}pke,~F.~K., Hillebrandt, W., \& Niemeyer, J.~C. 2004a, A\&A, 420, 411 
\bibitem{Rea042} R\"{o}pke,~F.~K., Hillebrandt, W., \& Niemeyer, J.~C. 2004b, A\&A, 421, 783 
\bibitem{RH05a} R\"{o}pke,~F.~K, \& Hillebrandt,~W. 2005a, A\&A, 429, L29
\bibitem{RH05b} R\"{o}pke,~F.~K, \& Hillebrandt,~W. 2005b, A\&A, 431, 635
\bibitem{RH05c} R\"{o}pke,~F.~K, \& Hillebrandt,~W. 2005c, in preparation 
\bibitem{SFD98} Schlegel,~D.~J., Finkbeiner,~D.~P., \& Davis,~M. 1998, ApJ, 500, 525
\bibitem{Sch04} Schmidt,~W., Hillebrandt,~W., \& Niemeyer,~J.~C. 2005, in preparation
\bibitem{SLKea04} Sollerman. J., Lindahl J., Kozma, C., 
\etal 
2004,  A\&A, 428, 555
\bibitem{SF54} Spencer, L.~V. \& Fano, U. 1954, Phys.Rev, 93, 1172 
\bibitem[Spyromilio et al. (2004)]{Spyro} Spyromilio,~J., Gilmozzi,~R., Sollerman,~J., Leibundgut,~B., Fransson,~C., \& Cuby,~J.-G., 2004, A\&A, 426 547 
\bibitem{SMBH04} Stehle,~M., Mazzali,~P.~A., Benetti,~S., \& Hillebrandt,~W., 2004, astro-ph/0409342
\bibitem{SUN99} Suntzeff,~N.~B., Phillips,~M.~M., Covarrubias,~R., Navarrete,~M., P\'{e}rez,~J.~J., Guerra,~A., Acevedo,~M.~T., Doyle,~L.~R., Harrison,~T., Kane,~S., Long,~K.~S., Maza,~J., Miller,~S., Piatti,~A.~E., Clari\'{a},~J.~J., Ahumada,~A.~V., Pritzl,~B., Winkler,~P.~F., 1999, AJ, 117, 1175
\bibitem{TNY86} Thielemann, F.-K., Nomoto, K., \& Yokoi, K. 1986 A\&A,
158, 17 
\bibitem{TNH96} Thielemann, F.-K., Nomoto, K., \& Hashimoto, M. 1996, ApJ, 460, 408
\bibitem{T03} Thomas, R.~C. 2003, to appear in proc.
'3-D Signatures in Stellar Explosions', Austin, Texas,
astro-ph/0310619
\bibitem{TBBNLF03} Thomas, R.~C., Branch D., Baron, E., Nomoto, K., Li, W., \&
Filippenko, A.~V. 2004, ApJ, 601, 1019 
\bibitem{TKBB02} Thomas, R.~C., Kasen, D., Branch, D., \& Baron, E. 2002, ApJ, 567, 1037
\bibitem{THRT04} Travaglio, C., Hillebrandt, W., Reinecke, M., \& 
Thielemann, F.-K. 2004, A\&A, 425, 1029
\bibitem{vR62} van Regemorter, H. 1962, ApJ, 136, 906 
\bibitem{WWK04} Woosley,~S.~E., Wunsch,~S., \& Kuhlen,~M. 2004, ApJ, 607, 921
\bibitem{YNST92} Yamaoka,~H., Nomoto,~K., Shigeyama,~T., \& Thielemann,~F. 1992, ApJ, 393, L55
\end{thebibliography}
\end{document}